\documentclass[prl,aps,twocolumn,superscriptaddress]{revtex4}
\usepackage{graphicx}

\begin{document}

\title{Anomalous Spin Dephasing in (110) GaAs Quantum Wells:\\
Anisotropy and Intersubband Effects}

\author{S. D\"{o}hrmann}
\affiliation{Institut f\"{u}r Festk\"{o}rperphysik,
Universit\"{a}t Hannover, Appelstra\ss e 2, D-30167 Hannover,
Germany}
\author{D.~H\"{a}gele}
\affiliation{Institut f\"{u}r Festk\"{o}rperphysik,
Universit\"{a}t Hannover, Appelstra\ss e 2, D-30167 Hannover,
Germany}
\author{J. Rudolph} \affiliation{Institut f\"{u}r
Festk\"{o}rperphysik, Universit\"{a}t Hannover, Appelstra\ss e 2,
D-30167 Hannover, Germany}
\author{D. Schuh}
\affiliation{Walter Schottky Institut, Technische Universit\"{a}t
M\"{u}nchen, Am Coulombwall, D-85748 Garching, Germany }
\author{M. Bichler}
\affiliation{Walter Schottky Institut, Technische Universit\"{a}t
M\"{u}nchen, Am Coulombwall, D-85748 Garching, Germany }
\author{M.~Oestreich}
\affiliation{Institut f\"{u}r Festk\"{o}rperphysik,
Universit\"{a}t Hannover, Appelstra\ss e 2, D-30167 Hannover,
Germany}

\begin{abstract}
A strong anisotropy of electron spin decoherence is observed in
GaAs/(AlGa)As quantum wells grown on (110) oriented substrate. The
spin lifetime  of spins perpendicular to the growth direction is
about one order of magnitude shorter compared to spins along
(110). The spin lifetimes of both spin orientations decrease
monotonically above a temperature of 80 and 120~K, respectively.
The decrease is very surprising for spins along (110) direction
and cannot be explained by the usual Dyakonov Perel dephasing
mechanism. A novel spin dephasing mechanism is put forward that is
based on scattering of electrons between different quantum well
subbands.

\end{abstract}
\pacs{78.66.Fd, 73.50.-h, 85.75.-d}

\maketitle


 The electron spin in
semiconductors has recently become  a focus of intense research in
the context of spinelectronics or spintronics. This new kind of
electronics aims to utilize spin for devices with unprecedented
properties \cite{wolfscience01,
 imamogluPRL99, rudolphAPL03}. A
prime condition for the development of potential applications is
the understanding of spin decoherence, {i.e.}~the loss of spin
memory, in semiconductor structures \cite{synonymsNOTE}. The main
reason for spin decoherence at room temperature is the intrinsic
spin splitting of the conduction band, which occurs in all binary
semiconductors. The spin splitting, which acts as an effective
magnetic field, depends on the electron's momentum and is the
basis for the Dyakonov-Perel (DP) spin relaxation mechanism
\cite{dyakonovSPSS72,opticalorientationBOOK84}. Semiconductor
heterostructures are in this context of particular interest since
spin splitting in conduction and valence band can be controlled
via dimensionality and orientation of crystal axes
\cite{kainzPRB03}. Ohno et al. observed very long electron spin
decoherence times at room temperature in GaAs quantum wells (QWs)
grown on (110) oriented substrates that exceeded the coherence
times in usual (100) grown QWs by more than one order of magnitude
\cite{ohnoPRL99,dyakonovSPS86}. However, slow spin dephasing in
(110) QWs had been demonstrated only for electron spins oriented
along the crystal growth direction. The dynamics of in-plane spin
was left unexplored.


Starting point for the theoretical description of the spin
dynamics in (110) QWs is the Dresselhaus-Hamilton for binary
semiconductors
\begin{equation}
 H_{\rm spin} = \Gamma \sum_i \sigma_i k_i(k_{i+1}^2 -
 k_{i+2}^2), \label{bulkspinsplitting}
\end{equation}
where $i=x,y,z$ are the principal crystal axes with $i+3
\rightarrow i$, $\Gamma$ is the spin-orbit coefficient for the
bulk semiconductor, and $\sigma_i$ are the Pauli spin matrices
\cite{hassenkamPRB97}. Comparing eq. (\ref{bulkspinsplitting})
with the spin Hamilton for a free electron in a magnetic field ($H
= \frac{1}{2} \sum_i \mu_{\rm B} \sigma_i B_i$) one easily
recognizes that random scattering of electrons leads to an
effective $k$ dependent random magnetic field with components in
$x$, $y$, and $z$ direction. This random magnetic field destroys
the average spin orientation of an ensemble of electrons by
rotating individual spins in different directions. The DP effect
increases in bulk semiconductors with temperature due to
occupation of higher $k$-states with larger spin splittings
despite a motional narrowing effect at higher temperatures (spin
lifetime $\tau_s$ is inversely proportional to momentum scattering
time $\tau_p^*$). In (110) QWs, however, the spin splitting
(effective magnetic field)
\begin{equation}
  H_{\rm DP} = -\Gamma \sigma_z k_x \left[\frac{1}{2} \langle{k_z^2}\rangle -
  \frac{1}{2}(k_x^2 - 2 k_y^2)\right],
\end{equation}
is suppressed in the plane of the QW via averaging of $H_{\rm
spin}$ along the confinement direction $z\parallel [110]$ (Ref.
\cite{hassenkamPRB97}). Here, $\langle{k_z^2}\rangle = \int
|\nabla \psi_z|^2\,dz$, $\psi_z$ denotes the $z$-part of the
electron wavefunction in the lowest subband,  $x\parallel
[1\overline{1}0]$, and $y
\parallel [001]$. An effective magnetic field exists only along the
growth direction $z$. Electron spins oriented along $z$ are
therefore not affected by the DP mechanism. In contrast, electrons
with a spin component along $x$ or $y$ will randomly precess
around the effective field, giving rise to spin dephasing
anisotropy with dephasing rates $\gamma_x = \gamma_y \gg
\gamma_z$.

In the first part of the paper we report on the observation of
such a huge dephasing anisotropy with fast dephasing of electron
spins oriented in the plane of a (110) QW and long dephasing times
for spins along growth direction.
The sample under investigation is a symmetrically modulation doped
GaAs/Al$_{0.4}$Ga$_{0.6}$As multiple quantum well (MQW) with 10
wells of 20~nm thickness and an electron density of $n = 1.1\cdot
10^{11}$~cm$^{-2}$ per well. The sample is mounted in a magnet
cryostat with variable temperature. The magnetic field lies in the
plane of the MQW along the $y$ direction. Spin dephasing rates are
determined by means of time and polarization resolved optical
photoluminescence (PL) measurements. Spin oriented electrons are
optically created in the sample along the growth direction by
circularly polarized 100~fs pulses  from a modelocked Ti:Sapphire
laser with a repetition rate of $80$~MHz. The excitation energy is
at 1.675~eV, i.e. about  200~meV  above the PL energy. After
excitation the carrier momentum distribution rapidly thermalizes
by emission of phonons and scattering with other carriers. Holes
rapidly lose their spin orientation due to strong valence band
mixing and $k$ dependent spin splitting.  The polarized
luminescence is spectrally and temporally resolved by a
synchroscan streakcamera with two-dimensional readout which
provides a resolution of $1$~nm and $20$~ps, respectively.
Polarization is selected by a liquid crystal retarder and a
polarizer \cite{PlNOTE}. The optical selection rules yield a
proportionality between the degree of circular PL polarization and
the  degree of spin orientation, with opposite sign for heavy hole
(hh) and light hole (lh) transitions
\cite{opticalorientationBOOK84}. Photoluminescence lifetimes
increase with temperature from 500 ps at 6~K to 8~ns between 200~K
and 300~K, which is sufficiently long for a precise determination
of spin relaxation times.

\begin{figure}
   \centering
   \includegraphics[width=7cm]{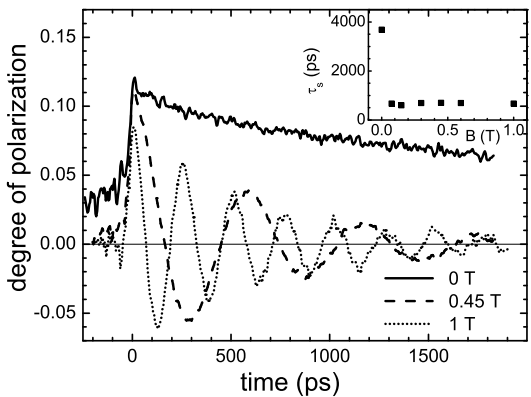}
   \caption{Degree of polarization of the time resolved PL for magnetic fields of $0$ (solid line),
    $0.45$ (dashed line) and $1$~T (dotted line) measured at a temperature of
    200~K. Inset: Dependence of measured spin lifetime $\tau_s$ on magnetic fields.
    }
    \label{fig1}
\end{figure}

 Figure 1 depicts the transient decay of
circular polarization after excitation of about $2\cdot 10^{10}$
carriers per cm$^2$ at a sample temperature of 200~K. For $B = 0$,
the average spin direction is always oriented along the $z$
direction where the DP mechanism does not apply. We find a long
spin lifetime $\tau_z = \gamma_z^{-1} = 3700$~ps as reported by
Ohno \cite{ohnoPRL99}. For $B>0$ we observe spin precession around
the field which leads according to the optical selection rules to
a periodic change between $\sigma^-$ and $\sigma^+$ polarized
luminescence in $z$ direction \cite{heberlePRL94}. The magnetic
field forces the spin after a quarter precession period in the
plane of the QW, where the effective fluctuating magnetic field
rapidly randomizes the spin orientation of the ensemble of
electrons. We measure for the precessing spin an average dephasing
rate $(\gamma_z + \gamma_x)/2 = 1/670$~ps$^{-1}$ which is by a
factor of five faster than for spin in $z$-direction. The
dephasing rate $\gamma_x = 1/370$~ps$^{-1}$ is even ten times
larger than $\gamma_z$. Such a strong anisotropy in a
semiconductor structure has never been reported before
\cite{dyakonovNOTE}. The rapid dephasing rates $\gamma_x$ and
$\gamma_y$ have direct negative implications for devices where the
stability of spins with arbitrary orientation is important, like
in qubits for quantum computation.  We measure the magnetic field
dependence of the spin dephasing time to verify that the fast spin
dephasing is not a new dephasing effect caused by the external
magnetic field or inhomogeneous broadening of the precession
frequency (inset of Fig. \ref{fig1}). The almost constant spin
lifetime for $B>0$ clearly excludes a direct influence of the
magnetic field on spin dephasing. We note that this spin dephasing
for an electron precessing around an in-plane magnetic field is
fully understood in the DP one-electron picture with an effective
field in $z$ direction and does not require an additional
"many-body inhomogeneous broadening effect" as claimed by Wu
\cite{wuSSC02}.

\begin{figure}
   \centering
   \includegraphics[width=7cm]{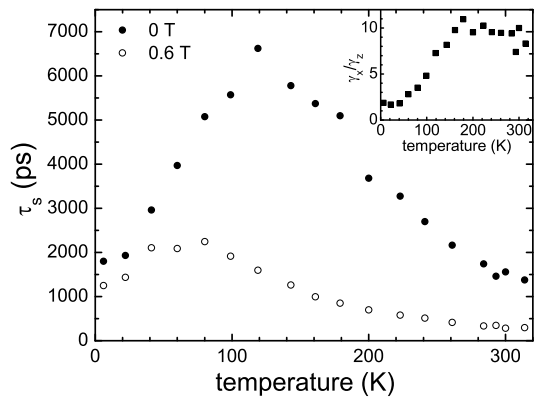}
   \caption{Temperature dependence of spin lifetime $\tau_s$ for $B = 0$ (closed circles) and $B = 0.6$~T (open circles).
   Corresponding temperature dependence of spin relaxation anisotropy
   $\gamma_x/\gamma_z$ (inset).
    }
    \label{fig2}
\end{figure}

Next, we systematically investigate the spin lifetimes for
temperatures between 6 and 314~K keeping the excitation power of
the laser fixed. Figure \ref{fig2} depicts the temperature
dependence of the spin lifetimes at $B = 0.6$~T and $B=0$ for the
full temperature range. Two features, the weak dephasing
anisotropy at low temperatures  ($\gamma_x/\gamma_z$, see inset)
and the decreasing spin lifetime for $B=0$ at temperatures above
$120$~K will be discussed in the following. At low temperatures
the spin dephasing is governed by the Bir-Aronov-Pikus (BAP)
mechanism where the randomly oriented hole spins act via spin
interaction with electron spins like a fluctuating effective
magnetic field in all directions \cite{birJETP76}. The main
contribution of the effective magnetic field arises from the 1s
exciton where the electron hole overlap is at maximum
\cite{fuPRB99}. In a quantum well the $z$ component of the
effective magnetic field is dominant,  which explains the
anisotropy $\gamma_x/\gamma_z > 1$  at low temperatures
\cite{blackwoodPRB94}.  Higher temperatures break the strong
spatial correlation of electrons and holes which leads to the
observed increase in spin lifetime up to $80$~K. Another clear
hint for the dominance of dephasing via the BAP mechanism at low
temperatures is found in the time resolved PL transients, where we
observed a retarded sub-exponential decay of the degree of  spin
polarization. Here the BAP mechanism decreases in time due to loss
of holes by optical recombination (not shown)
\cite{holedephasingnote}.

For $B = 0.6$~T and $T > 80$~K, the DP mechanism becomes stronger
than the BAP mechanism and  leads to the expected reduced spin
lifetime $[(\gamma_x+\gamma_z)/2]^{-1}$. In contrast, the decrease
of spin lifetime $\gamma_z^{-1}$ for $B=0$ and $T
> 120$~K is very surprising, since all subbands strictly exhibit
no in plane component of the effective magnetic field. Occupation
of higher subbands (about 13\% for the first excited subband at
250 K  compared to the lowest band) at elevated temperatures
should also not lead to spin dephasing. This can be seen by
transforming eq. (\ref{bulkspinsplitting}) into the (110)
coordinate system
\begin{equation}
H_{\rm spin} = \frac{1}{2} \hbar \left[ \sigma_x \Omega_x +
\sigma_y \Omega_y + \sigma_z \Omega_z \right] ,
\label{110spinsplitting}
\end{equation}
where $(\Omega_x,\Omega_y,\Omega_z) = \Gamma/\hbar ( -k_x^2 k_z -
2 k_y^2 k_z + k_z^3,\,  4 k_x k_y k_z,\, k_x^3 - 2 k_x k_y^2 - k_x
k_z^2)$. Since $\langle k_z \rangle = \langle k_z^3 \rangle = 0$
for all subbands, all in plane magnetic field components
$\Omega_x$ and $\Omega_y$ vanish. A simple application of the DP
dephasing formula $\gamma_z = \langle \Omega_x^2 + \Omega_y^2
\rangle \tau_p^* = 0 $ yields no spin dephasing
\cite{opticalorientationBOOK84}.

In the following we resolve the puzzle by noting that scattering
of electrons between subbands along with spin orbit coupling
$H_{\rm spin}$ constitutes a new spin relaxation mechanism.
Existing theories treat spin dephasing only within the lowest
subband \cite{wuPRB00,lauPRB01,glazovJETP02,pullerPRB03}. Here, we
sketch the principle idea for intersubband spin relaxation (ISR)
and give a first estimate of its effectiveness. In the case
$H_{\rm spin}=0$ a simultaneous intersubband and spin flip
transition like $|\vec{k},0,\uparrow \rangle \longrightarrow
|\vec{k'},1,\downarrow \rangle$ is forbidden. However, for $H_{\rm
spin}\neq 0$ the corresponding transition $|\vec{k},0,\uparrow
\rangle + \epsilon_1 |\vec{k},1,\downarrow \rangle + ...
\longrightarrow |\vec{k'},1,\downarrow \rangle + \epsilon_2
|\vec{k'},0,\uparrow \rangle + ... $ is modified according to
first order perturbation theory which mixes zero order terms with
first order terms possessing opposite spin. Therefore spin flip
transitions become allowed with rates that are $\alpha =
|\epsilon_1|^2+|\epsilon_2|^2$ slower than the intersubband
transition rate $\tau_{\rm IB}^{-1}$, i.e. $\tau_s \approx
\alpha^{-1} \tau_{\rm IB}$. We obtain the first order coefficient
$\epsilon_1(k_x,k_y) = \langle \vec{k},0,\uparrow|H_{\rm spin
}|\vec{k},1,\downarrow\rangle/E_{\rm ISG}$ using a spin splitting
constant $\Gamma = 19.55$~eV{\AA}$^{3}$ \cite{winklerbook}, an
intersubband gap of $E_{\rm ISG} = 31$~meV and approximating the
$z$-wavefunctions with that of a well with infinite barriers. We
find a spin lifetime $\tau_s = \langle \alpha \rangle^{-1}
\tau_{\rm IB} = 1.7$~ns, for an intersubband scattering rate
$\tau_{\rm IB}=250$~fs and a distribution of electrons at 300~K
with averaged $\alpha$ (effective mass $m^* = 0.0665 m_0$). The
estimate is in good agreement with our measurements. ISR depends
strongly on the occupation of higher subbands and will be much
weaker in narrow quantum wells where intersubband splitting is
large. This is in accord with recent measurements on an 8.3~nm
thick n-doped GaAs QW that showed spin lifetimes of up to 20 ns at
room temperature \cite{adachiPHYSICAE01}. We note that our new
dephasing mechanism may also explain a result from Ohno where he
found an increased spin lifetime in higher mobility (110) samples
at room temperature ($\tau_s \propto \tau^*_p,\tau_{\rm IB}$
respectively), which he could not explain with the usual DP result
($\tau_s \propto (\tau^*_p)^{-1}$) \cite{ohnoPRL99}.

Decreasing spin lifetimes for elevated temperatures have also been
reported in (110) grown samples with external electric field. This
field induces a Rashba term ($\Omega_x, \Omega_y \neq 0$) which
leads via the DP mechanism to spin relaxation \cite{karimovPRL03}.
In our sample, the Rashba terms should be very weak or absent
since asymmetry of the QW or a built in electric field in growth
direction is suppressed by symmetric growth and symmetric doping.
We moreover found no experimental evidence for the presence of a
Rashba contribution which should reveal itself in spin lifetimes
which strongly depend on the orientation of a tilted external
magnetic field $(B_x \cos(\theta),B_y \sin(\theta),B_z)$ if $B_z
\neq 0$ \cite{rashbaNOTE2}. And even if some Rashba related spin
relaxation was present in our sample, the above estimate shows
that the novel ISR mechanism contributes substantially.

In the last paragraph we argue and experimentally verify that
anisotropic spin dephasing modifies the usual linear relation
$\omega = g \mu_{\rm B} B / \hbar$  between magnetic field $B$ and
measured frequency $\omega$ of spin oscillations, where $\mu_{\rm
B}$ is Bohr's magneton and $g$ is the constant Land\'{e} g-factor.
The time dependent degrees of spin polarization $s_x$ and $s_z$
for spins which precess around a magnetic field oriented along $y$
are given by
\begin{equation}
\frac{\partial}{\partial t} \left( \begin{array}{c} s_x \\ s_z
\end{array} \right) = -\left( \begin{array}{cc} \gamma_x & -\omega \\ \omega & \gamma_z
\end{array} \right)\left( \begin{array}{c} s_x \\ s_z
\end{array} \right),
\end{equation}
where $\gamma_x$ and $\gamma_z$ are the relevant components of the
spin dephasing tensor. The solution for $s_z(t)$ which is measured
in our experiments is
\begin{equation}
s_z(t) = s_0 \frac{e^{-(\gamma_x + \gamma_z)t/2}}{\cos \varphi}
\cos(\omega_{\rm a} t - \varphi), \label{pseudog}
\end{equation}
where $\tan\,\varphi = \frac{\gamma_x - \gamma_z}{2 \omega_{\rm a}
}$, $\omega_{\rm a} = \sqrt{\omega^2 - (\gamma_x-\gamma_z)^2/4}$,
and $s_z(0) = s_0$ is the initial degree of spin polarization. The
effective spin oscillation frequency $\omega_{\rm a}$ is directly
affected by the dephasing anisotropy $\gamma_x-\gamma_z$. The
slowing down of the spin oscillations for anisotropic spin
dephasing has a pictorial explanation. The magnetic field forces
the initial spin orientation into the $x$ direction. The fast
$x$-dephasing reduces the $x$ component, whereas the slow
$z$-dephasing leaves the $z$-component nearly unchanged. Reducing
the $x$ component drives the spin orientation back towards the $z$
axis effectively reducing the precession frequency. The result is
a delayed rotation of the spin into the plane of the QW giving
rise to a phase shift $\varphi$ of the spin oscillations. After
crossing the plane of the QW, the spin is accelerated by the same
effect towards the $-z$ direction. On average the spin precession
frequency is reduced. Figure 3 (a) depicts the measured modified
oscillation frequency $\omega_{\rm a}$  for magnetic fields
between 0.075 and 1~T along with a fit of $\omega_{\rm a}(B)$. The
value of $\gamma_x - \gamma_z = $1/380~ps$^{-1}$ obtained from the
fit compares very well with the expected value of
$1/410$~ps$^{-1}$ extracted from the data in Fig. \ref{fig2}. The
most significant influence of anisotropic spin dephasing is found
 at
magnetic fields below 0.1~T. Figure 3(b) - 3(d) shows the measured
time resolved degree of spin polarization along with the
theoretical curves including anisotropy [Eqn. (\ref{pseudog})] and
without anisotropy, i.e. $s_z(t) = s_0 e^{-(\gamma_x +
\gamma_z)t/2} \cos(\omega t)$, where we used the initial degree of
polarization $s_0$ as the only fit parameter. The values for the
parameters used for calculating the transients were obtained as
follows: $\gamma_z$ by fitting the transient at $B~=~0$ and $g$
and $\gamma_x$ by fitting the transients at elevated magnetic
fields ($0.075$ to $1$~T), assuming validity of anisotropic spin
dephasing. While the transient at $B = 0.6$~T is matched by both
theoretical curves [Fig. 3 (b)], the transients for $B = 0.075$~T
and  $B = 0.025$~T can only be explained by including anisotropic
spin dephasing. We note that for the case of $|\omega| < |\gamma_x
- \gamma_z|/2$, i.e. $B < 0.05$~T, spin oscillations are
completely suppressed.

\begin{figure}
   \centering
   \includegraphics[width=7cm]{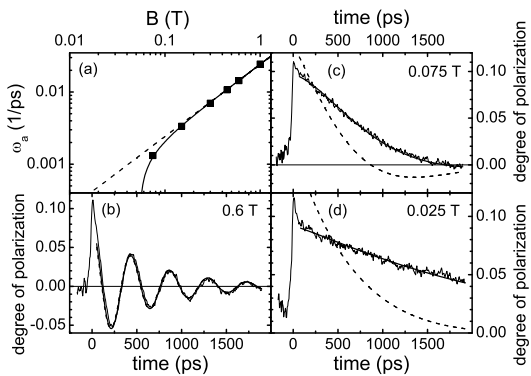}
   \caption{(a) Measured spin oscillation frequency $\omega_{\rm a}$  vs. magnetic field (closed
   squares) at $T = 200$~K.
    The fitting curve includes effects of anisotropic spin dephasing (black
    line). The dashed line depicts for comparison the spin
    oscillation frequency for isotropic spin dephasing.
     (b)-(d) Transients of the degree of polarization of time resolved PL for
    magnetic fields of $0.6$, $0.075$, and $0.025$~T. The black lines
    are fits including effects of anisotropic spin dephasing. The dashed lines show fits assuming no anisotropy.}
    \label{fig3}
\end{figure}

In conclusion we investigated anisotropic spin dephasing in (110)
QWs from 6~K up to room temperature. Spin memory at room
temperature lasts for times as long as nanoseconds only if the
spin is parallel or antiparallel to the growth direction. Spin
coherence in the plane of the QW is rapidly lost by the DP
mechanism. These results have an immediate impact on design
considerations for spintronic devices. We further found that
intersubband scattering constitutes a new spin dephasing mechanism
that is necessary to explain reduced spin lifetimes at elevated
temperatures for electron spins along (110). We expect that this
new mechanism will stimulate the development of a more general
theory of spin dephasing in QWs including the influence of higher
subbands. Finally we demonstrated the modification and total
suppression of spin oscillations via anisotropic spin dephasing.

We gratefully acknowledge most helpful discussions with R. Winkler
and financial support by BMBF and DFG (German Science Foundation).



\begin{thebibliography}{10}
\bibitem{synonymsNOTE}
Through all the paper the terms spin decoherence, dephasing, and
relaxation are
  used as synonyms.
\bibitem{PlNOTE}
The width of the luminescence peak increase with temperature from
3.5 to 12 nm.
  For temperatures above 100~K we used the low energy side of the PL spectrum
  to determine spin polarization (peak to lower energetic half of the hh
  transition). A possible contribution of the broadened light hole transition
  above 50~K can lower the degree of PL polarization but does not obscure spin
  relaxation data, because the ratio of hh and lh emission exhibits no
  transient change. Holes are at elevated temperatures almost instantly in
  thermal equilibrium.
\bibitem{holedephasingnote}
The spin dephasing times for $B~=~0$ in Fig. \ref{fig2} at low
temperatures are
  deduced from the best exponential fit in the time interval between $0$ and
  $1.8$ ns.
\end{thebibliography}
\end{document}